# Deciphering the dynamics of the light-induced phase transition in VO$_2$

**Authors:** S. Mandal[1], P. Kumar[1], H. Y. Kim[1], Z. Pi[1], J. Xu[2], D. Chen[2], D. Kazenwadel[3], P. Baum[3], S. Meng[2,*] and E. Goulielmakis[1,*]

**Affiliations:**

[1]Institut für Physik, Universität Rostock; Albert-Einstein-Straße 23–24, 18059 Rostock, Germany.

[2]Institute of Physics, Chinese Academy of Sciences, Beijing 100190, China

[3]Universität Konstanz, Fachbereich Physik, 78464 Konstanz, Germany.

[*]Authors to whom correspondence shall be addressed. Email: e.goulielmakis@uni-rostock.de, smeng@iphy.ac.cn

**Vanadium dioxide (VO$_2$) is central in the study of ultrafast photoinduced insulator-to-metal phase transitions in strongly correlated materials, and a primary candidate for next-generation light-driven devices. However, the physical mechanism underlying its phase transition remains unresolved. Here, we use single-cycle light transients to perform phase-resolved ultrafast spectroscopy on VO$_2$ crystals. Our experiments reveal two processes: a structural transformation from the insulating monoclinic M1-VO$_2$ phase to the excited metallic rutile R$^*$-VO$_2$ phase, followed by electron thermalization and relaxation dynamics intrinsic to the newly formed excited metallic phase. By disentangling the structural and electronic contributions we retrieve the anharmonic V–V dimer bond oscillations that underlie the conversion of the monoclinic lattice into its metallic counterpart on its natural timescale. Our work lays the foundation for real-time tracking of insulator-to-metal phase transitions in correlated matter and opens prospects for controlling these phenomena on their native time scales.**

**Main Text:**

Exploration and control of fundamental properties of solids such as the electrical conductivity, magnetic state and optical response on ultrafast time scales is central for the development of electronic/optoelectronic devices operating at unprecedented speeds. Strongly-correlated transition metal oxides—such as vanadium-based compounds that undergo insulator-to-metal phase transitions (ITM) through electronic or structural pathways—have long been regarded as prime systems for exploring these possibilities. Among these, vanadium dioxide ($VO_2$) stands out as a noteworthy candidate. When heated above a critical temperature $T_C=68°C$, $VO_2$ undergoes a first order phase transition (Fig. 1A) from an insulating monoclinic phase M1-$VO_2$ to a conductive rutile phase R-$VO_2$ (*1*, *2*), accompanied by an increase of the electrical conductivity by more than three orders of magnitude. Importantly, this phase transition can also be triggered by the absorption of light (*3*, *4*). While it is well understood that light-induced phase transition is initiated by the electronic photoexcitation of carriers in the monoclinic phase M1, the precise unfolding of electronic/structural dynamics and their role in the transient variation of the optical and electronic properties of the system remain elusive.

Ultrafast optical pump–probe techniques tracking the reflection (*4–7*) and/or transmission (*8*), photoelectron emission (*9*) and XUV absorption (*10*) of $VO_2$ specimen upon their optical excitation offer high temporal resolution and sensitivity to both electron and atomic dynamics, have so far yielded a range of often contradicting conclusions about both the timescale and consequently the nature of the ITM phase transition. While early pioneering studies on the femtosecond time scale emphasized (*4*, *5*) a primarily structural mechanism whose time scale is linked to the period of the

V–V vibrational mode in $VO_2$ (*11*), later investigations supported the notion of a pure electronic, Mott localization mechanism occurring nearly instantaneously (*8, 9*). Yet, these approaches cannot unambiguously resolve the transient structure of the material upon its optical excitation, leaving room for ambiguity. In contrast, time-resolved x-ray (*4, 12*), electron diffraction (*13–15*) and x-ray scattering (*16, 17*) probing techniques can provide structure-resolved tracking of the excited system dynamics, yielding important insights on the process, yet they remain insensitive to the electronic response and lack the temporal resolution typically attained by optical probing approaches. Thus far, these techniques have suggested that the ITM phase transition unfolds on a slower timescale (~100 fs) compared to those indicated by state-of-the art optical approaches (~26 fs) (*8, 10*) leaving behind a gap in the understanding of the intriguing phenomenon. Resolving these issues calls for approaches that can fill the essential gap, i.e., by combining the ultimate temporal resolution of optical techniques and the capability of phase-resolving the transition in real-time. Here we demonstrate that ultrafast spectroscopy with single-cycle light transients spectrally extending in the visible and adjacent ranges can bridge this gap.

In our experiments (Fig.1), we performed transient reflectivity measurements on precisely characterized, macroscopic, monocrystalline M1-$VO_2$ specimen (see supplementary text, Fig. S2) using single-cycle light transients (~2.4 fs) generated in a light-field synthesizer (*18, 19*). Measurements were performed in a versatile, vacuum-based experimental setup (Fig. S1 for details) which enables a non-collinear, all optical, pump-probe geometry (Fig. S1B), in tandem with in-situ tracking of the instantaneous, pulse field-waveform using attosecond streaking (*20*). The latter allows precise evaluation of the temporal resolution and measurement fidelity in the full spectral range of the probe pulse (see supplementary text). The pump pulse fluence was set to ~21 mJ/cm$^2$, well above the previously identified threshold, (*6, 14, 21*) for triggering an efficient insulator-to-metal phase transition in VO₂. In contrast, the probe pulse— more than an order of magnitude weaker— ensured that no phase transition was induced at negative pump-probe delays.

Time-resolved, differential reflectivity $S_{M1}(\lambda,\tau) = \frac{\Delta R(\lambda,\tau)}{R_{M1}(\lambda)}$ data in M1-VO₂ recorded at room temperature as a function of the delay $\tau$ of the super-octave (400 nm to 1000 nm) light transient is shown in Fig. 1B. Apart from the marked areas (white arrows in Fig. 1B) where the weak pedestal (< 2%) of the driving pulse gives rise to narrowband protrusions (see supplementary text, Fig. S3) extending over tens or hundreds of femtoseconds, the temporal resolution of our experiments was better than ~4 fs in all spectral areas critical for this study.

Fig. 1C-E display the spectrally averaged $S_{M1}(\lambda,\tau)$ in three ranges of the spectrogram in Fig. 1B: the near infrared (NIR) range (870 nm to 940 nm) (red points, Fig. 1C), the yellow range (540 nm to 640 nm) (orange points, Fig. 1D) and the blue range (420 nm to 480 nm) (blue points, Fig. 1E). While the decline of $S_{M1}(\lambda,\tau)$ in all these ranges aligns well with the observations of previous narrowband studies (*5, 6, 21, 22*), the decay-time constants $\Gamma^{-1}$ separately evaluated (dashed black curves in Fig. 1C-E) in these ranges exhibit notable differences. In the visible range the evaluated decay constant of $\Gamma_Y^{-1} = 102\,fs \pm 2\,fs$ (Fig. 1C) aligns well with the results of previous experiments (*4, 5, 21, 23*), yet in the near infrared (red points in Fig. 1C), the transient reflectivity drops approximately three times faster in comparison to that in the yellow range, with a decay constant of $\Gamma_{NIR}^{-1} = 31\,fs \pm 2\,fs$. The latter result is more compatible with the conclusions of other sets of experiments in the mid-infrared (MIR) (*8*) and the extreme ultraviolet (EUV) (*10*) ranges which have highlighted a considerably faster-time scale of the ITM phase transition in VO₂. The hitherto less examined blue range (Fig. 1E) displays a slightly but statistically significant shorter decay-time constant $\Gamma_B^{-1} = 98\,fs \pm 2\,fs$ compared to the yellow range.

Also notable in our measurements is the rapid decrease of differential reflectivity versus delay exhibiting a conspicuous multistep-like structure, well supported by the experimental accuracy (error bars in the inset of Fig. 1D). This structure is apparent over the entire spectrum of the probe pulse but with the relative amplitude between consecutive steps being non-uniform (Fig.1C-E).

The wavelength dependence of the decay times of $S_{M1}(\lambda,\tau)$ observed in our experiments sets a new stage of inquiry for the mechanisms underlying the phase transition in $VO_2$. Since the inherent timescale of the process should be independent of wavelength, the disparate decay time scales suggest the presence of hitherto unrecognized dynamics that directly influence the transient optical reflectivity of the system. Our observations therefore call into question – the long-standing assumption that the decay/rise in transient reflectivity/transmission maps one-to-one with the reduction/increase of the relative populations among the insulating and metallic phases in ultrafast experiments (*5, 6, 8, 21*). Identifying and disentangling these dynamics is thus essential for a thorough understanding of the ITM phase transition.

To pinpoint whether the apparent differences in the decay time scales observed in our experiments are linked to the transition from M1-$VO_2$ to its excited rutile counterpart, we conducted a second set of experiments designed to isolate and probe the intrinsic dynamics of R*-$VO_2$. To this end, we raised the temperature of the M1-$VO_2$ to 80°C to ensure that the sample entirely resided in its R-$VO_2$ phase. Under the same experimental conditions and geometry as discussed earlier, our pump pulse now promoted R-$VO_2$ to its excited state R*-$VO_2$ while the probe pulse tracked the ensuing reflectivity dynamics.

Fig. 2A presents time-resolved differential reflectivity $S_R(\lambda,\tau) = \frac{\Delta R(\lambda,\tau)}{R_R(\lambda)}$ data in metallic R-$VO_2$ over the bandwidth of our single-cycle pulses next to those recorded at room temperature (M1-

VO$_2$) (Fig. 2B) for facilitating their comparison. Inspection of the differential reflectivity data in R-VO$_2$, averaged over the identical spectral intervals (red lines in Fig. 2C-D) as the data of our previous experiment, highlights the characteristic optical response established in ultrafast experiments of metals under optical excitation (*24*). More specifically, $S_R(\lambda, \tau)$ exhibits a fast build-up (~25 fs) at early times — typically attributed to electron thermalization upon optical excitation — followed by a substantially slower relaxation (>100 fs) (*25–28*). Notably and in contrast to previous studies limited to narrower spectral ranges (*22*), the ultrabroad spectral coverage of our light transients exposes an essential characteristic of metallic R*-VO$_2$: the sign reversal of the differential reflectivity around ~834 nm (arising due to a shift in the plasma frequency (*29, 30*) of R-VO$_2$) as shown in Fig. 2A and Fig. 2C-D (red dots) (see also supplementary text, Fig. S4).

To understand how this generic characteristic of free-carrier plasma in metals and semiconductors (*24, 31*) serves as a spectroscopic fingerprint of R*-VO$_2$ in our main experiments, it is useful to consider how its formation would affect the decay rates of the transient reflectivity in different parts of the spectrum. If the optical excitation of M1-VO$_2$ drives the lattice to the excited rutile state, the electron heating-relaxation dynamics of the R*-VO$_2$ population should introduce (i) a transient negative contribution to the differential reflectivity at energies lower than the inflection point at ~834 nm (red dots in Fig. 2C) and (ii) a transient positive contribution in the yellow range (Fig. 2D red dots). This results in (i) the shortening of the transient reflectivity dynamics in the NIR range (Fig. 2C blue dots) and (ii) its extension in the yellow range (Fig. 2D blue dots). A second inspection of the data (Fig. 1C-D) reveals that they are precisely in agreement with the experimental observation and as such offer strong evidence for the formation of R*-VO$_2$ virtually at any delay between pump and probe pulses. This conclusion is also compatible with the relatively faster dynamics observed in the blue range (Fig. 2E) which, in this case, can be attributed to the interplay between the expected elongation in time scale and the negative baseline shift induced by $R^*$-$VO_2$ at long delays, leading to a weak but apparent shortening of the decay time relative to that in the yellow range.

To disentangle the structural and electron thermalization dynamics, we consider the transient reflectivity $S_{M1}(\lambda, t)$ to be composed of two main parts: (i) a component directly proportional to time-dependent rutile population $N_R(t)$ and (ii) a time evolving thermalization-relaxation response that builds up with the population rate of $R^*$-$VO_2$ as $\int_{-\infty}^{t} \frac{dN_R(t')}{dt'} S_R(\lambda, t - t')dt'$, where $S_R(\lambda, t)$ is the intrinsic instantaneous response function of rutile upon excitation. $S_R(\lambda, t)$ in Fig. 2A is a reasonable approximation for this intrinsic instantaneous response function based on the auxiliary experiments (Fig. 2), as the ultrashort laser pulse excites R-$VO_2$ on a timescale much shorter (~2.4 fs) than its electron thermalization-relaxation dynamics (tens of fs).

Thereby, we can express $S_{M1}(\lambda, t)$ as:

$$S_{M1}(\lambda, t) = [a(\lambda) - 1]N_R(t) + a(\lambda) \int_{-\infty}^{t} \frac{dN_R(t')}{dt'} S_R(\lambda, t - t')dt' \qquad (1)$$

Here, $a(\lambda)$ is a wavelength dependent coefficient which describes the reflectivity ratio between pure R-$VO_2$ and M1-$VO_2$. Moreover, as $S_{M1}(\lambda, t)$ and $S_R(\lambda, t)$ are sampled over a broad spectral range, the dynamics of the system can be described via a large set (> 900) of independent equations of the form of Eq. (1) describing the identical underlying population dynamics (see supplementary text). Solving the resulting system of equations through regularized, parametric optimization allows the disentanglement of the two main components of transient reflectivity of the system (see supplementary text).

Fig. 3A displays the population dynamics of the monoclinic phase $N_{M1}(t) = 1 - N_R(t)$ as retrieved based on the above methodology. Notably, the multistep decline of the transient reflectivity observed in the original data (Fig. 1) retains its statistical significance. A closer look into the multistep structure is possible by evaluating the corresponding population conversion rate (Fig. 3B, green curve) $-dN_{M1}/dt$ in Fig. 3A, which highlights that the initial M1-VO$_2$ phase population is converted into its excited metallic counterpart R$^*$-VO$_2$ in merely two consecutive bursts of decreasing amplitude and time span. The stepwise drop of M1-VO$_2$ population is also supported by the close inspection of the retrieved component of the transient reflectivity associated with the electron thermalization-relaxation dynamics (Fig. 3C) in the excited rutile for a representative wavelength of 580 nm. These data further suggest that electron thermalization-relaxation dynamics are triggered by a sequence of excitation steps of the R$^*$-VO$_2$ phase.

We now proceed to elucidate the link between the derived time-dependent populations ($N_{M1}(t)$ and $N_R(t)$) and the underlying lattice dynamics of the system. To this end we assume that after photoexcitation, the system evolves along a single lattice coordinate Q(t) that captures the ultrafast V-V dimer motion. Operationally, we identify Q(t) with the absolute V-V dimerization length difference $d(t) = |d_{V-V}^{long}(t) - d_{V-V}^{short}(t)|$, where $d_{V-V}^{long}$ and $d_{V-V}^{short}$ are the dimerization lengths of the long and short V-V pairs respectively. In this picture the tendency for metallization of the system is governed by the instantaneous value of Q(t). More specifically, we model the forward (M1-VO$_2$ → R-VO$_2$) $\Lambda_{M1 \to R}(t)$ and backward (R-VO$_2$ → M1-VO$_2$) $\Lambda_{R \to M1}(t)$ rates as exponential functions of Q(t), consistent with the exponential dependence of electron coupling in tight binding approaches (*32, 33*) (see supplementary text). Within this framework, the system population is partitioned into two phases or states, M1-VO$_2$ and R$^*$-VO$_2$, and evolves as a function of time according to the rates that depend on the instantaneous lattice distortion. The master equation can be written with the populations of M1-VO$_2$ and R-VO$_2$ as:

$$\frac{dN_{M1}(t)}{d\tau} = -\Lambda_{M1 \to R}(t)N_{M1}(t) + \Lambda_{R \to M1}(t)N_R(t)$$

A fitting of the experimentally derived population data based on this premise allows the evaluation of the dimerization length difference dynamics $d(t)$ as shown in Fig. 3D.

While the dynamics observed are apparently not compatible with the simplified scenario of light induced motion under the fundamental, single-frequency, phonon V-V mode, they are supportive of earlier theoretical predictions that assigned the ITM phase transition to be the result of the highly anharmonic motion of the V-V dimers (*16, 34–36*). To further support these conclusions, we conducted detailed real time time-dependent density functional theory (TDDFT) simulations in M1-VO$_2$ under similar conditions prevalent in our experiments, including the detailed (Fig. S3A) driving pulse waveform as recorded in our setup. Fig. 4A shows simulated V-V pair length dynamics for both the long $d_{V-V}^{long}$ (yellow curve) and short $d_{V-V}^{short}$ (purple curve) dimerization amplitudes respectively. Fig. 4B (brown curve) shows their absolute difference $d(t)$ and allows direct comparison with the experimentally derived dynamics (crimson dashed curve). The comparison reveals an excellent agreement between experiments and simulations for both the rapid initial anharmonic distortion of V-V pair length as well as the subsequent lattice oscillation. The experimentally observed faster decay of $d(t)$ compared to our simulations may be anticipated given that the theoretical model does not necessarily capture all possible dephasing pathways, in particular, those arising from the thermalization and relaxation processes of the excited metallic phase.

The above experiments and simulations allow us to draw a transparent physical picture underlying the photoinduced ITM phase transition in $VO_2$ via direct tracing of the relevant dynamics. The short-pulse photoexcitation of M1-$VO_2$ modifies the interatomic potential to trigger highly anharmonic atomic dynamics in the electronic excited system that results in the shortening of the long V-V dimer and the corresponding elongation of the shorter one to convert the system to its rutile counterpart. The system is progressively converted to rutile (Fig. 3A), reaching maximum conversion rates each time the anharmonic oscillators tend to equalize the V-V dimer distances (Fig. 3B and Fig. 4A). The electronically excited rutile formed by the above process in turn undergoes electron thermalization and relaxation dynamics (Fig. 3C) that co-evolve with the rapid-population conversion.

By extending the repertoire of modern ultrafast solid-state spectroscopy of correlated materials to include synthesized single-cycle light transients, we uncover previously hidden, strongly anharmonic lattice dynamics and identify them as the key rate-limiting factor in the phase transition. More significantly, our study reveals that the ultrafast response of $VO_2$—as a candidate for ultrafast optoelectronic switching—is governed by the interplay of structural and electronic degrees of freedom. This coexistence determines the switching speed across different spectral regions. Consequently, our findings provide a blueprint for engineering ultrafast phase-change functionality in $VO_2$. Tailoring light fields to control lattice trajectories may enable solid-state switching at frequencies extending into the tens of terahertz. More broadly, our phase-resolved spectroscopic approach introduces a powerful new toolkit for mapping and controlling non-equilibrium phase transitions in complex materials.

**Acknowledgments:**

This work was supported by the European Research Council (project number - 101098243) and Deutsche Forschungsgemeinschaft (project numbers - 441234705 and 437567992). We thank Dr. Rostyslav Lesyuk from Prof. Dr. Christian Klinke's group and Annika Bergmann-Iwe from Prof. Dr. Tobias Korn's group at the University of Rostock for their invaluable help in the characterization of the sample used in the experiments.

**Author contributions:**

SM, HYK and EG designed the experiments. SM, PK, HYK and ZP performed experiments. PK, HYK, ShM, SM, JX, DC conducted simulations. PB and DK developed the macroscopic $VO_2$ crystals. EG conceived and supervised the experiments. All coauthors contributed to the preparation of the manuscript.

**Competing interests:** Authors declare that they have no competing interests.

**Supplementary Materials**

Supplementary Text

Figs. S1 to S5

References (37–72)

# Figures

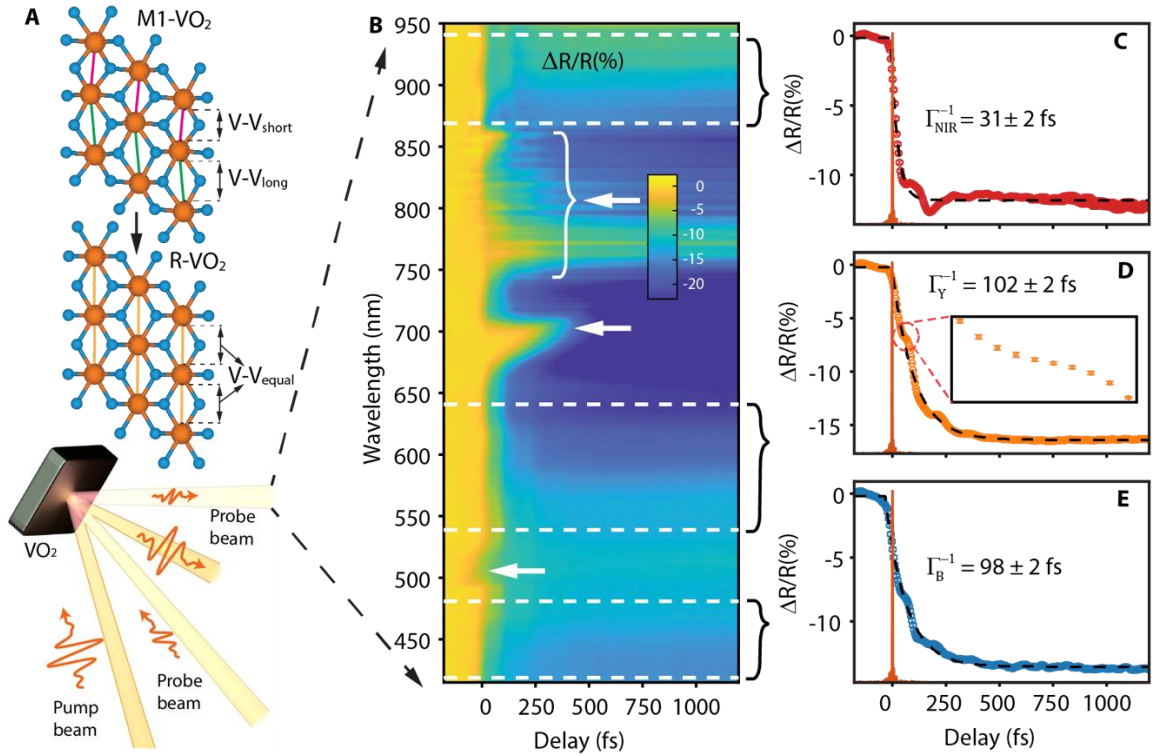

**Fig. 1. Single-cycle probing of the insulator-to-metal phase transition in VO₂.** (**A**) (Top) Crystal structure of monoclinic (M1) and rutile (R) VO$_2$. (Bottom) Schematic of the experimental pump-probe setup. (**B**) Broadband spectrogram representing the variation in differential reflectivity (in percentage) across the spectral range of 415 nm – 950 nm. The white arrows mark spectral protrusions arising from weak pulse pedestals at the corresponding wavelengths ranges. Spectrally averaged transient differential reflectivity in three sub-regions (870 nm – 940 nm, 540 nm – 640 nm and 420 nm – 480 nm) as indicated by the white dashed lines. The black curly braces indicate the corresponding averaged differential reflectivity in (**C**), (**D**), and (**E**) as red, orange and blue circular points respectively. The error bars in (C), (D) and (E) denote the standard error of the aforementioned spectral averages from all measurement samples. The inset in (D) zooms on a portion of the trace to pinpoint the step-like structure. Black dashed lines in (C), (D), and (E)

represent fits to an exponential decay. The sharp curve in (C), (D), and (E) (deep orange) at the delay t=0 shows the intensity profile of the single-cycle light transient.

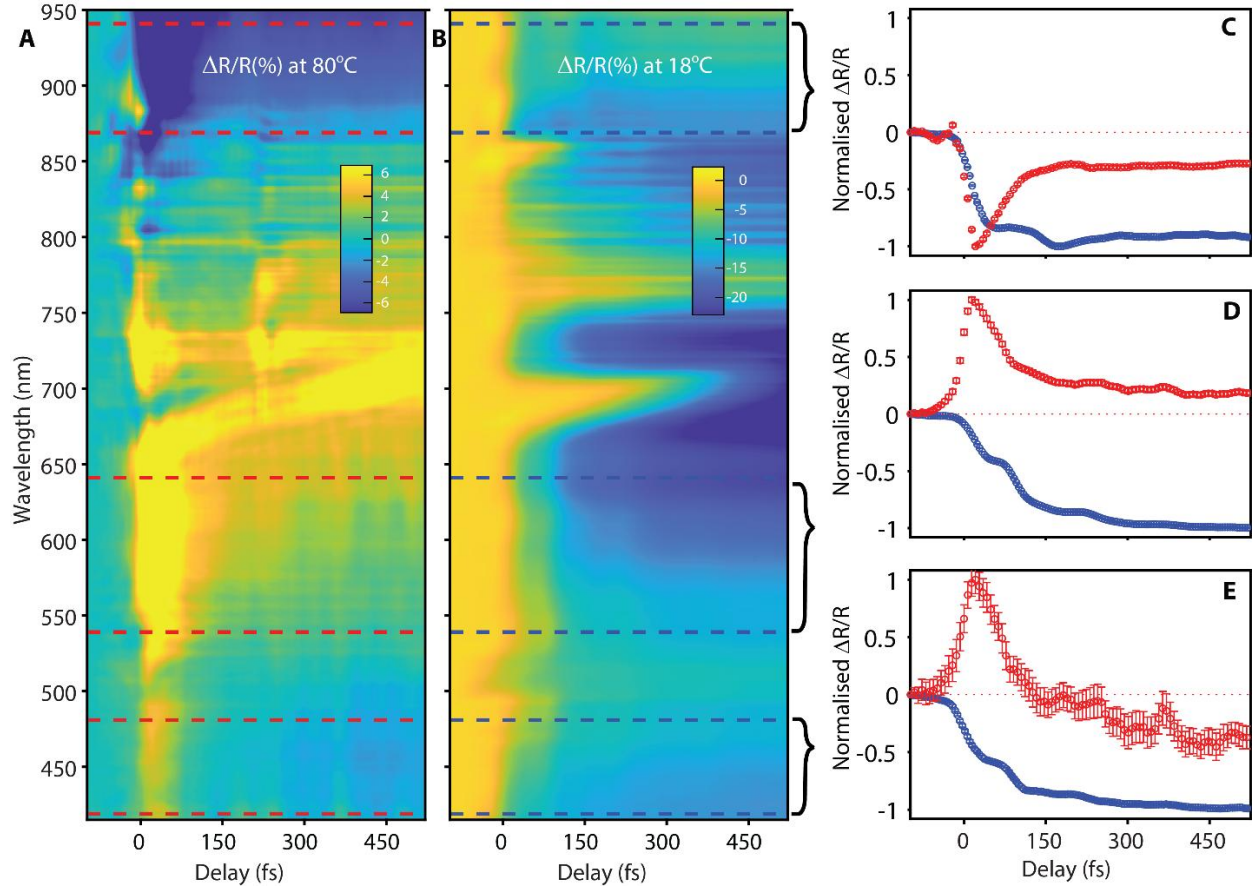

**Fig. 2. Single-cycle transient reflection dynamics in R-VO$_2$ and M1-VO$_2$ in comparison.**

(**A**) Transient differential reflectivity of R-VO$_2$ (**B**) The data of Fig. 2A for facilitating comparison. The normalized, spectrally-averaged transient differential reflectivity in three spectral ranges (870 nm – 940 nm, 540 nm – 640 nm and 420 nm – 480 nm), indicated by red dashed lines in A and blue dashed lines in B, are depicted in (**C**), (**D**) and (**E**). The error bars in (C), (D) and (E) denote the standard error of the aforementioned spectral averages from all measurement samples and the red dotted lines mark the zero level.

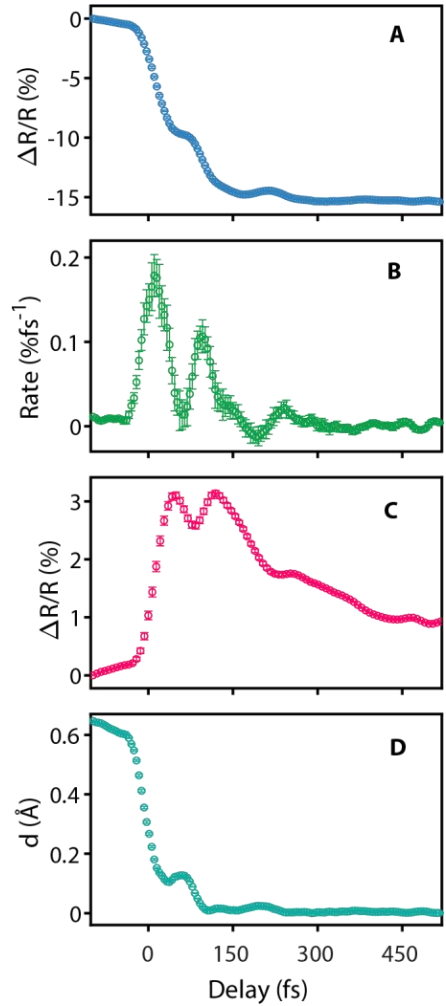

**Fig. 3. Structural and electronic dynamics of the ITM phase transition in VO$_2$.** (**A**) Retrieved population dynamics $N_{M1}(t)$, where the error bars denote the unbiased deviation from the principal value. (**B**) Population rate $-dN_{M1}/dt$ based on the data in (A), whose error bars are propagated from those in (A). (**C**) Retrieved electron thermalization and relaxation dynamics of R$^*$-VO$_2$, whose error bars are propagated from those in (A) and the standard errors of the measured spectrogram samples. (**D**) Retrieved V-V dimerization length difference dynamics, with its error bars derived from the minimization process as discussed in the supplementary text.

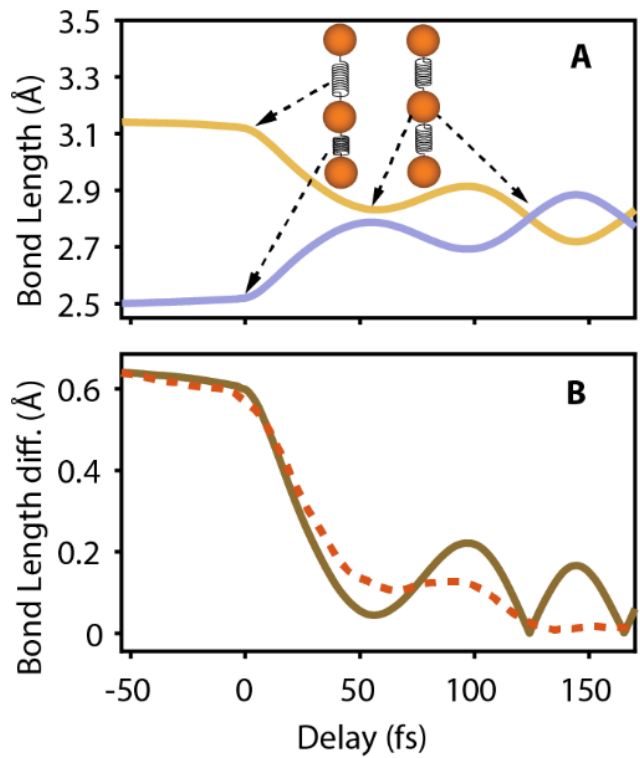

**Fig. 4. Tracking the anharmonic structural dynamics in VO$_2$ (Theory vs Experiment)**

(**A**) Simulated V-V long (yellow solid line) and short (purple solid line) dimerization length dynamics following single-cycle pulse photoexcitation of VO$_2$. Ball and spring models illustrate the V-V dimers at representative instances. (**B**) Absolute difference between the V-V long and short dimers as evaluated from the data in (A) (brown solid line). The experimentally retrieved dimerization length difference dynamics for comparison (crimson dashed line).